# Intrinsic Properties of Stoichiometric LaOFeP

T. M. McQueen[1], M. Regulacio[1], A. J. Williams[1], Q. Huang[2], J. W. Lynn[2], Y. S. Hor[1], D.V. West[1], M. A. Green[2,3], and R. J. Cava[1]

[1]Department of Chemistry, Princeton University Princeton NJ 08544
[2]NIST Center for Neutron Research, National Institute of Standards and Technology, Gaithersburg MD 20899
[3]Department of Materials Science and Engineering, University of Maryland, College Park, Maryland 20742-2115, USA

**Abstract**

DC and ac magnetization, resistivity, specific heat, and neutron diffraction data reveal that stoichiometric LaOFeP is metallic and non-superconducting above $T = 0.35\ K$, with $\gamma = 12.5\ \frac{mJ}{mol\cdot K^2}$. Neutron diffraction data at room temperature and $T = 10\ K$ are well described by the stoichiometric, tetragonal ZrCuSiAs structure and show no signs of structural distortions or long range magnetic ordering, to an estimated detectability limit of 0.07 $\mu_B/Fe$. We propose a model, based on the shape of the iron-pnictide tetrahedron, that explains the differences between LaOFeP and LaOFeAs, the parent compound of the recently discovered high-Tc oxyarsenides, which, in contrast, shows both structural and spin density wave (SDW) transitions.

**Introduction**

The compounds LnOFeX (Ln = La…Gd, X =P,As) were first synthesized in 1995 (X = P)[1] and 2000 (X = As)[2]. However, it was not until 2006 that the first report of superconductivity in this class of compounds, in LaOFeP, was published.[3] Since then, a range of transition temperatures have been reported in fluorine doped (LnO$_{1-x}$F$_x$FeX) and oxygen deficient (LnO$_{1-x}$FeX) variants, including $T_c$'s greater than $T = 50\ K$ in SmO$_{1-x}$F$_x$FeAs.[4-24] Thus this class of superconductors has the highest transition temperatures known except for the cuprates. The origin of the remarkably high $T_c$'s has been the subject of considerable debate. One interesting observation is that while

LaOFeAs is metallic but non-superconducting[25], LaOFeP has been reported as a superconductor.[3, 26, 27] The reported $T_c$'s in LaOFeP have varied from T = 3.1 $K^3$ to 7 K[27], and appear to depend on the sample form and synthesis conditions.[26] However, to date there has been no complete study of LaOFeP. In particular, neither resistivity nor susceptibility measurements (used in the above cited works on LaOFeP) are conclusive proof of bulk superconductivity, as both can be sensitive to impurities. Instead, the presence or absence of an anomaly in the specific heat is a more reliable indicator of bulk properties, including superconductivity. The original aim of our work was to investigate the superconductivity in pure LaOFeP. Instead we have found, though an exhaustive set of measurements, that stoichiometric LaOFeP is not a superconductor above $T = 0.35\ K$. Additionally, we find that it is a metal with no long range magnetic ordering or structural distortions above $T = 10\ K$. These properties make LaOFeP closer to LaOFeAs than previously thought. We also provide a chemical explanation for the difference in the behavior of LaOFeP compared to LaOFeAs.

**Experimental**

Polycrystalline samples of LaOFeP were synthesized in multiple steps. First, LaP was synthesized by reacting fresh La shavings and dry P powder in an alumina crucible in a sealed, evacuated silica ampoule; the ampoule was immediately heated to 400 °C, then ramped at 20 °C/hr to 800 °C. The temperature was held for 20 hr, and then the sample was furnace-cooled. Next, a stoichiometric ~2.5 g mixture of LaP, Fe, and $Fe_3O_4$ was ground together and pressed into a pellet. This pellet was placed with 5 % excess P in an alumina crucible with a tight-fitting alumina cap. This crucible was placed on quartz shards inside a quartz tube. A second, smaller, alumina crucible containing ~80 mg of impure LaOFeP from a previous run was also placed in the quartz tube, which was then pumped down and sealed under vacuum. The sample was ramped to 1200 °C at 180 °C/hr, held at temperature for 48 hr, and then cooled at 180 °C/hr to room temperature. After that the sample was removed, reground, repressed, and heated with an additional 5% excess P as before. This was repeated a third time to achieve phase purity. In each repetition, the pellets of LaOFeP came out clean, with no visible undesired reactivity, even on the surface. The alumina crucibles and quartz tubes were also clean and undamaged. In contrast, the impure LaOFeP getterer on top came out visibly different in color; x-ray diffraction confirmed the presence of silicate formation. In further processing the material was protected from air as it was found to decompose with extended exposure to moisture.

DC magnetization measurements between $T = 1.8\,K$ and $T = 300\,K$ were performed on a quantum design MPMS magnetometer with applied fields of $\mu_0 H = 0.0005\,T$ and $\mu_0 H = 1\,T$. AC magnetization measurements were performed on a quantum design PPMS with a dc field of $\mu_0 H_{dc} = 0.0005\,T$ and an ac field of $\mu_0 H_{ac} = 0.0003\,T$ at $f = 1\,kHz$. Resistivity and specific heat measurements were done on polycrystalline pellets between $T = 0.35\,K$ and $T = 300\,K$ in a quantum design PPMS equipped with a $^3$He refrigerator. Thermopower measurements were done using a custom-built helium probe-head and MMR technologies electronics.

High resolution neutron powder diffraction (NPD) data were collected using the BT-1 high-resolution powder diffractometer at the NIST Center for Neutron Research, employing Cu (311) monochromator to produce a monochromatic neutron beam of wavelength 1.5403 $Å$. Collimators with horizontal divergences of 15′, 20′, and 7′ full width at half maximum were used before and after the monochromator, and after the sample, respectively. The intensities were measured in steps of 0.05° in the 2θ range 3-168°. The structure analysis was performed using the program

GSAS with EXPGUI.[28, 29] The neutron scattering amplitudes used in the refinements were 0.827, 0.954, 0.581 and 0.513 ($\times 10^{-12}$ cm) for La, Fe, O, and P, respectively. To investigate possible magnetic ordering, additional data were collected on the high intensity/coarse resolution BT-7 spectrometer with a pyrolytic graphite monochromator and filter using a wavelength of 2.44 Å, and a position sensitive detector in diffraction mode.

**Results**

Figure 1 shows neutron powder diffraction patterns collected at room temperature and $T = 10K$, along with the fits from Rietveld refinements. The patterns are well described by the tetragonal phase LaOFeP, and this structure is shown in Figure 2(a). It consists of P-Fe$_2$-P layers of edge-sharing Fe-P tetrahedra separated by La-O$_2$-La-type sheets. The iron ions form two dimensional square nets, where the Fe-Fe distance is 2.80 Å at room temperature. Table 1 summarizes the results from the structure fits. Since previous results in the As case suggest that fluorine doping or oxygen non-stoichiometry induce superconductivity, we allowed the occupancies of the La, O, and P sites to vary in the $T = 298 K$ refinement. Within error, all occupancies are equal to one (column 1, Table 1). As a further test of the stoichiometry, the thermal parameters were held fixed and all four occupancies were refined (column 2, Table 1). Again, within error all occupancies are equal to one and thus were fixed at unity for the final refinements (columns 3 and 4, Table 1). The quality of the fits is excellent. We included minor impurity phases of 0.6 % La$_2$O$_3$ and 2 % FeP in the final refinements; the LaOFeP is stoichiometric, as indicated by the free refinement of the occupancies. The inset of Figure 1(a) shows a comparison of the (220) reflection at $T = 298 K$ and 10 K. There is no sign of peak broadening or splitting that would indicate a structural transition similar to that observed in LaOFeAs.[8, 14] Furthermore, there is no observable difference in the residuals between $T = 298 K$ and 10 K, implying a lack of long range magnetic order. As an additional check for weak magnetic super-reflections, low angle, high count rate neutron diffraction data were collected at $T = 7\,K$, 100 $K$, 200 $K$, and 320 $K$. All observed peaks are indexed and well-fit by the nuclear structure of LaOFeP. Thus we observe no long range magnetic order in LaOFeP. The detectability limit indicates that any ordered moment, if the magnetic structure is analogous to that in LaOFeAs,[14] would have to be less than 0.07 $\mu_B$/Fe. Thus we conclude that, above $T = 10\,K$, LaOFeP shows no structural distortion or long range magnetic ordering.

In addition to being metallic, the magnetic susceptibility (Figure 3) data are essentially flat from $T = 1.8\,K$ to $T = 300\,K$, showing contributions only from Pauli paramagnetism and Landau diamagnetism (corrections for core diamagnetism and the sample holder were applied). This is in contrast to an impure specimen containing traces of $Fe_2P$ which shows pronounced magnetic behavior. There is a slight upturn in the susceptibility of the pure sample between $T = 200\,K$ and 250 $K$ that is attributable to the impurity FeP (Curie temperature 215 $K$[30]), and a further small upturn below $T = 20\,K$ that may be due to either a paramagnetic impurity or the proposed spin fluctuations (see below). However, these features are small and indicate the lack of magnetism in the susceptibility of a pure sample. When combined with the neutron diffraction data that show the absence of long range magnetic order (see above), this implies that LaOFeP is non-magnetic above $T = 1.8\,K$. For comparison to previous literature reports, we also show the dc susceptibility measured under a low field, $\mu_0 H = 0.0005\,T$, (top inset) as well as an ac measurement scaled per gram of sample (bottom inset). Both show negligible responses to the applied field, with no downturn at low temperatures. This indicates that stoichiometric LaOFeP is not superconducting above $T = 1.8\,K$.

To check for superconductivity at lower temperatures, resistivity measurements were done from $T = 0.35\ K$ to $T = 300\ K$ as shown in Figure 4(a). The data are normalized to the room temperature value ($\rho_{300} = 1.51\ m\Omega cm$) as the measured resistivity on a polycrystalline sample is often sensitive to grain boundary and surface effects, and higher than the intrinsic values. The resistivity decreases to $\rho_{20} = 0.61\ m\Omega cm$ at $T = 20\ K$. Furthermore, the thermopower (Figure 4(b)) is small and negative. Thus LaOFeP is an n-type metal. Importantly, there is no downturn in the resistivity even at the lowest temperatures (left half of inset Figure 4(a)), indicating that the stoichiometric sample of LaOFeP is not superconducting. However, the residual resistivity ratio (RRR) is 2.5, lower than that expected for a pure material free of defects and disorder, and there is a slight upturn in the resistivity below $T = 20\ K$ to $\rho_{0.35} = 0.64\ m\Omega cm$ at $T = 0.35\ K$. The most likely origin for both of these observations is scattering at grain boundaries in the polycrystalline pellet. However, these results could also be intrinsic to LaOFeP, and the result of weak localization or other effects. One indication that the upturn may be intrinsic is our observation of a small but positive magneto-resistance under an applied field of $\mu_0 H = 9\ T$ (right half of inset Figure 4(a)) that is greatest at the resistivity minimum. Positive magnetoresistances are unusual and not readily explained solely by grain boundary effects, and the fact that the magnetoresistance maximum occurs at the resistivity minimum suggests that the two are related. Furthermore, the thermopower data also show a minimum at a similar temperature, and the magnetization (see above) has a slight upturn. High quality single crystals would be helpful in determining the origin of these results. The present data show that LaOFeP is an n-type metal that is non-superconducting above $T = 0.35\ K$.

The nonmagnetic, metallic nature of LaOFeP is confirmed by specific heat measurements. Figure 5 shows the low temperature specific heat at $\mu_0 H = 0\,T$, 1 $T$, 3 $T$ and 9 $T$. At all three fields there is a sharp upturn at the lowest temperatures ($T < 1\,K$). A plot of $CT^2$ versus $T^3$ below $T = 1\,K$ (Figure 5 inset) is linear for all three cases. Thus the sharp increase can be described as the high temperature portion of a Schottky anomaly, with a contribution to the specific heat of $C_{shottky} = \frac{B}{T^2}$. This term is ascribed to the freezing out of nuclear spins of either $^{31}$P (S = 1/2) and/or $^{139}$La (S = 7/2). The lack of a comparable upturn in the case of CeOFeP,[31] and the presence of an upturn in the case of LaONiAs,[32] suggests that the origin is $^{139}$La. In addition to the Schottky anomaly there is a broad upturn in the $\mu_0 H = 0\,T$ data that starts at higher temperatures ($T \approx 5\,K$). This broad upturn is weakly enhanced at $\mu_0 H = 1\,T$, and suppressed by $\mu_0 H = 9\,T$. It is well-described by a logarithmic contribution to the specific heat of $C_{sf} = AT^3 \ln T$. Therefore the specific heat below $T = 10\,K$ was fit to the formula $C = \gamma T + \beta T^3 + AT^3 \ln T + \frac{B}{T^2}$, where γ, β, A, and B are refinable parameters. The fits are quite good, as shown in Figure 5, and the values of the parameters are given in Table 2. The fitted Sommerfeld coefficients (γ) are very close to what is expected[33,37] from the predicted density of states at the Fermi level, $\gamma_{calc} = 14.1\,\frac{mJ}{mol \cdot K^2}$, implying a negligible effective mass enhancement. This is also consistent with the measured susceptibility (Figure 3): $\chi_{meas} = 3.1 \cdot 10^{-9}\,\frac{m^3}{mol}$ ($= 2.4 \cdot 10^{-4}\,\frac{emu}{mol \cdot Oe}$) is close to the expected value[38], $\chi_{calc} = 2.4 \cdot 10^{-9}\,\frac{m^3}{mol}$ ($= 1.9 \cdot 10^{-4}\,\frac{emu}{mol \cdot Oe}$). Although it is impossible to separate the Pauli and Landau components of the measured susceptibility from our data, the Stoner enhancement factor, indicative of the degree of exchange-enhancement in the system, can be estimated as $S = \frac{3}{2} \cdot \frac{\chi_{meas}}{\chi_{calc}} = 1.9$, where the numerical prefactor converts the measured susceptibility to estimate the Pauli contribution.[39] This is small and, given the large predicted density of states at the Fermi level, would not be expected to cause an observable mass enhancement in γ. However, S > 1.0 implies exchange-enhanced behavior, and the presence of the logarithmic contribution to the specific heat is consistent with this picture. Taking the enhancement as arising from the presence of spin fluctuations (LaOFeP is predicted to be near a magnetic instability), the spin fluctuation temperature $T_{sf}$ and lattice contribution $\beta_3$ can be extracted from the fitted β and A values.[40] The values obtained are $T_{sf} = 14.3\ K$ and $\beta_3 = 0.199 \frac{mJ}{mol \cdot K^4}$. The small value of $T_{sf}$ is consistent with the observed suppression of the logarithmic term under high fields and the concomitant decrease in γ. It is also consistent with the Stoner enhancement factor estimated from the susceptibility data: using this $T_{sf}$, the measured γ, and the fitted coefficient A at $\mu_0 H = 1T$, the calculated[41] Stoner enhancement factor is S = 1.5, in reasonable agreement with S = 1.9 from the susceptibility data (especially given the number of assumptions). The value of $\beta_3$ corresponds to a Debye temperature of $\theta_D = 340\ K$, similar to what was found for LaONiAs. Thus the specific heat data of LaOFeP are consistent with LaOFeP being a non-magnetic metal with weak exchange enhancement from spin fluctuations. The origins of these fluctuations deserve further study, and could also explain the minimum in the resistivity and thermopower measurements.

**Discussion**

These results are substantially different than for the related compound LaOFeAs. Recent research has shown that LaOFeAs is a metal that undergoes a structural distortion at $T = 150\ K$ followed by the formation of a SDW at $T = 134\ K$.[8, 14] We see no evidence for either kind of transition in LaOFeP down to $T = 10\ K$ by neutron diffraction. This might be qualitatively explained by the fact that the Fe-Fe separation is larger in LaOFeAs than it is in LaOFeP (2.85 Å versus 2.80 $Å$), meaning that it should be closer to localized (magnetic) electron behavior. The Fe-Fe separation of undoped SmOFeAs is reported to be only 2.79 $Å$,[7] smaller than what we find for LaOFeP, but SmOFeAs exhibits the same peak in the resistivity that is seen at the structural transition in LaOFeAs near $T = 150\ K$.[7] This suggests that the in-plane metal-metal distance is not the critical factor in driving the structural distortion. Instead, we propose that the important structural feature is the shape of the iron-pnictide tetrahedron. Figure 2(b) shows a comparison of the Fe-pnictide tetrahedron in LaOFeP and LaOFeAs.[14] The Fe-X bond distance increases by 6 %, from 2.28 $Å$ (X = P) to 2.41 $Å$ (X = As), a consequence of the larger size of As relative to P. This change (0.13 $Å$) is close to what is expected from related compounds (e.g., the mean Fe-X bond distance increases by 0.13 $Å$ between FeP and FeAs). Although the bond lengths increase by 6 %, the in-plane Fe-Fe distance only increases by 2 %. This is consistent with the fact that the dimensions of the ionic La-$O_2$-La layer are expected to be fixed by the size of $La^{3+}$ and $O^{2-}$ ions. Since each lanthanum ion is also coordinated to four pnictide ions (Figure 2(a)), the in-plane dimensions of the X-$Fe_2$-X layer will be primarily determined by the La-$O_2$-La network, consistent with what is observed. Thus when going from P to As, the larger size of the pnictide results in a substantial expansion of the c-axis to obtain favorable Fe-As bond lengths. Thus the tetrahedron in LaOFeAs is less compressed than in LaOFeP (Figure 2(b)), with a top As-Fe-As bond angle of *113.0°* compared to a P-Fe-P angle of 120.6° (an ideal tetrahedron would have an angle of 109.5°). This decreases the Jahn-Teller splitting of the d-orbital derived bands (Figure 2(b)) and reduces the energy range spanned by d-derived states. In turn this means that intrasite electron correlations (Hubbard U) can drive the As system closer to localized, non-metallic behavior. This qualitative chemical explanation matches recent theoretical work which showed that LaOFeAs is close to opening a gap at the Fermi level due to electron-electron correlations.[34]

## Conclusion

In conclusion, our dc and ac magnetization, resistivity, specific heat, and neutron diffraction data show that stoichiometric LaOFeP is non-magnetic and non-superconducting above $T = 0.35\ K$. These results suggest that the superconductivity observed in the LaOFeP system may be due to either oxygen deficiency,[35] as has been reported in the case of LaOFeAs[16], or the presence of superconducting impurities such as $LaFe_4P_{12}$ ($T_c = 4.1\ K$)[36], lanthanum metal ($T_c = 6.9\ K$), or tin flux inclusions ($T_c = 3.1\ K$). In contrast to LaOFeAs, which shows both a structural and SDW transition, we find that LaOFeP is a normal metal with no magnetic behavior except a $T^3 \ln T$ contribution to the specific heat data at low temperatures that is attributable to spin fluctuations. We propose that the differences in the shape of the metal-pnictide tetrahedron between the P and As cases, due to the size difference of the pnictide and chemical pressure of the La-$O_2$-La framework, are responsible for the radically different properties observed.

## Acknowledgements

T. M. McQueen gratefully acknowledges support by the national science foundation graduate research fellowship program. The work at Princeton was supported by the Department of Energy, Division of Basic Energy Sciences, grant DE-FG02-98ER45706. Identification of commercial equipment in the text is not intended to imply recommendation or endorsement by the National Institute of Standards and Technology.

**Table 1**. Refined structural parameters at T = 298 K and 10 K. Space group P4/nmm (#129). Atomic positions: **La**: 2c (1/4,1/4,z), **Fe**: 2b (3/4,1/4,1/2), **P**: 2c (1/4,1/4,z), and **O**: 2a (3/4,1/4,0). In the first column, the formula was freely refined by fixing the occupancy of the iron site and letting all others vary. In the second column, the thermal parameters were held fixed and all occupancies were allowed to vary. The final refinements (columns 3 and 4) fixed all occupancies at unity. Anisotropic thermal parameters were used. The sample contained 0.7 % $La_2O_3$ and 2 % FeP. Lattice parameters are in units of Å, and thermal parameters are in units of $10^{-2}$ Å$^2$.

| LaOFeP | | T = 298 K (fixed iron n) | T = 298 K (fixed U) | T = 298 K (fixed n) | T = 10 K (fixed n) |
|---|---|---|---|---|---|
| | a | 3.96307(4) | 3.96306(4) | 3.96306(4) | 3.95667(4) |
| | c | 8.5087(1) | 8.5087(1) | 8.5087(1) | 8.4973(1) |
| La | z | 0.1488(2) | 0.1488(2) | 0.1487(2) | 0.1493(2) |
| | $U_{11}=U_{22}$ | 0.84(6) | 0.79 | 0.79(5) | 0.31(4) |
| | $U_{33}$ | 0.57(9) | 0.52 | 0.52(8) | 0.28(8) |
| | n | 1.02(1) | 1.004(4) | 1 | 1 |
| Fe | $U_{11}=U_{22}$ | 0.69(5) | 0.75 | 0.75(5) | 0.32(4) |
| | $U_{33}$ | 0.75(8) | 0.81 | 0.81(7) | 0.28(7) |
| | n | 1 | 0.995(4) | 1 | 1 |
| P | z | 0.6348(3) | 0.6347(3) | 0.6348(3) | 0.6354(3) |
| | $U_{11}=U_{22}$ | 0.8(1) | 0.80 | 0.80(8) | 0.56(7) |
| | $U_{33}$ | 0.7(1) | 0.7 | 0.7(1) | 0.2(1) |
| | n | 1.01(1) | 0.997(7) | 1 | 1 |
| O | $U_{11}=U_{22}$ | 0.80(9) | 0.73 | 0.73(7) | 0.58(6) |
| | $U_{33}$ | 0.9(1) | 0.8 | 0.8(1) | 0.4(1) |
| | n | 1.02(1) | 1.007(6) | 1 | 1 |
| | $\chi^2$ | 1.197 | 1.195 | 1.198 | 1.470 |
| | $R_{wp}$ | 6.19 % | 6.19 % | 6.20 % | 6.66 % |
| | $R_p$ | 4.93 % | 4.93 % | 4.94 % | 4.98 % |
| | $R(F^2)$ | 4.67 % | 4.73 % | 4.71 % | 4.64 % |

**Table 2**. Parameters extracted from fits of the low temperature specific heat of LaOFeP.

| Applied Field | $\gamma$ (mJ/mol*K$^2$) | $\beta$ (mJ/mol*K$^4$) | A (mJ/mol*K$^4$) | B (mJ*K/mol) |
|---|---|---|---|---|
| 0 T | 12.5(1) | -0.098(26) | 0.111(12) | 0.17(2) |
| 1 T | 13.1(1) | -0.117(20) | 0.120(8) | 0.13(2) |
| 3 T | 12.3(1) | 0.0004(2) | 0.072(4) | 0.10(1) |
| 9 T | 11.0(1) | 0.031(3) | 0.066(1) | 0.22(2) |

**Figure 1.** Neutron powder diffraction data with Rietveld fits at $T = 298\ K$ (a) and $T = 10\ K$ (b). The tick marks correspond to LaOFeP. The inset of (a) shows the lack of splitting in the (220) reflection on cooling.

**Figure 2.** (a) The structure of tetragonal LaOFeP consists of alternating La-$O_2$-La and P-$Fe_2$-P layers. One iron pnictide tetrahedron is shaded (purple), as are the planes forming a square antiprism coordination of a $La^{3+}$ ion. Ionic sizes were used for $La^{3+}$ and $O^{2-}$ whereas covalent sizes were used for Fe and P. (b) Extracted portions of the iron-pnictide layer for LaOFeP and LaOFeAs, showing both the square net of iron ions and the difference in the coordination tetrahedron in each case. Also shown is a simplified picture of how the less compressed tetrahedron in the arsenide case, which results in a smaller Jahn-Teller splitting ($\delta$), would be expected to decrease the total bandwidth of the d-orbital-derived bands (B) and make the arsenide closer to localized, non-metallic behavior.

**Figure 3.** DC magnetization data at $\mu_0 H = 1\ T$ (main panel) and $\mu_0 H = 0.0005\ T$ (top inset) show than LaOFeP is non-superconducting and non-magnetic above $T = 1.8\ K$. The dc magnetization on an impure sample containing 2% $Fe_2P$ is also plotted. For comparison to the literature, the bottom inset shows an ac magnetization measurement of this sample that shows no trace of superconductivity (the scale is 10x more sensitive than previous reports)[3, 26, 27].

**Figure 4.** (a) Resistivity measurements on a polycrystalline pellet show that LaOFeP is metallic and displays a slight positive magnetoresistance. The left inset shows the low temperature region confirming the non-superconducting nature of this sample. The right inset shows the observed resistivity minimum, possibly from spin fluctuations. (b) Thermopower data show that LaOFeP is an n-type metal. An upturn at low temperatures is also observed (inset).

**Figure 5.** Specific heat measurements show both a Schottky anomaly below 1 $K$ and $T^3 \ln T$ behavior responsible for the broad upturn around 5 $K$. The lines are fits to the data (see text). The $T^3 \ln T$ contribution is suppressed under an applied field, suggesting that its origin is spin fluctuations. The inset shows a fit to the $T < 1\ K$ data showing good agreement with a Schottky anomaly.

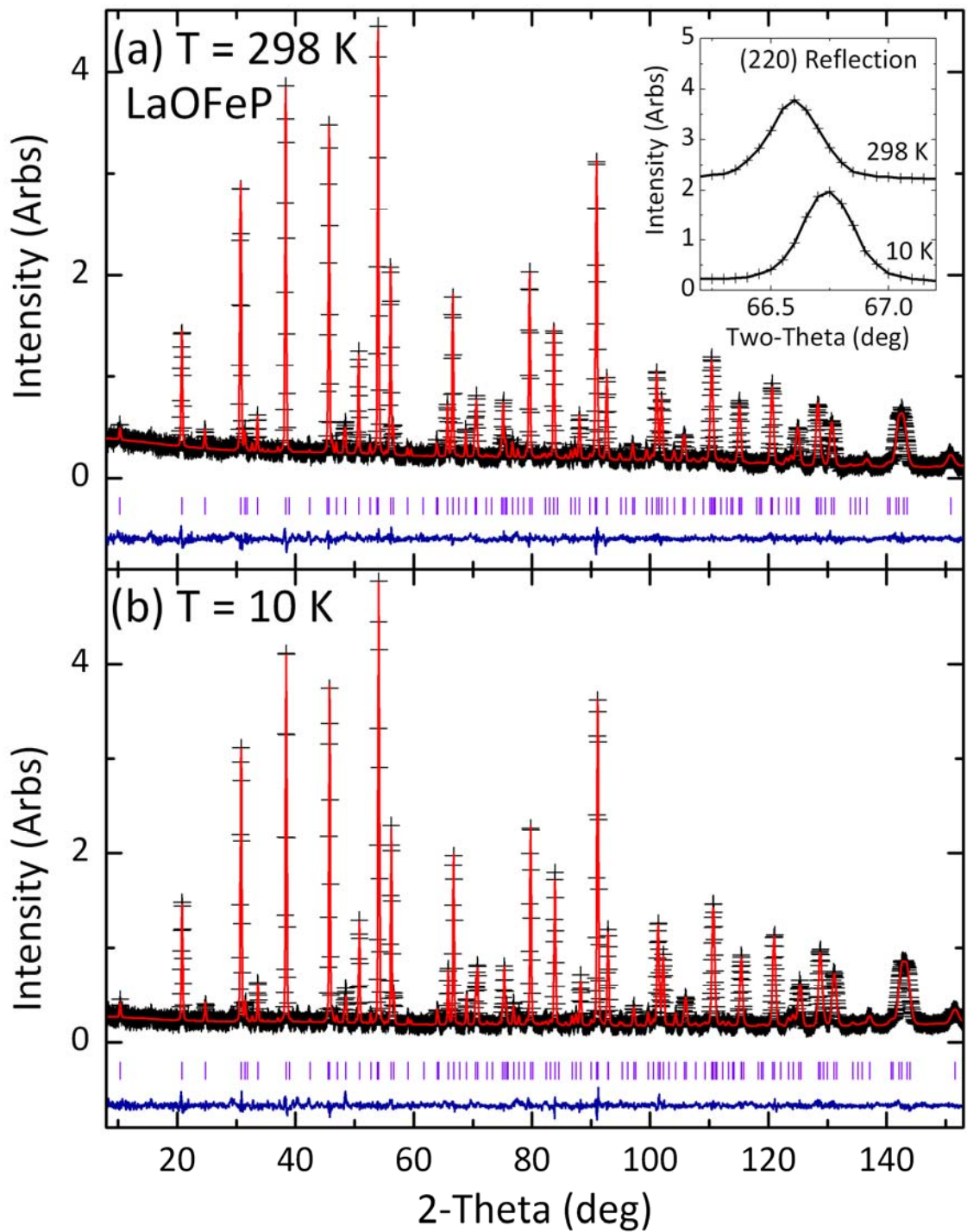

Figure 1

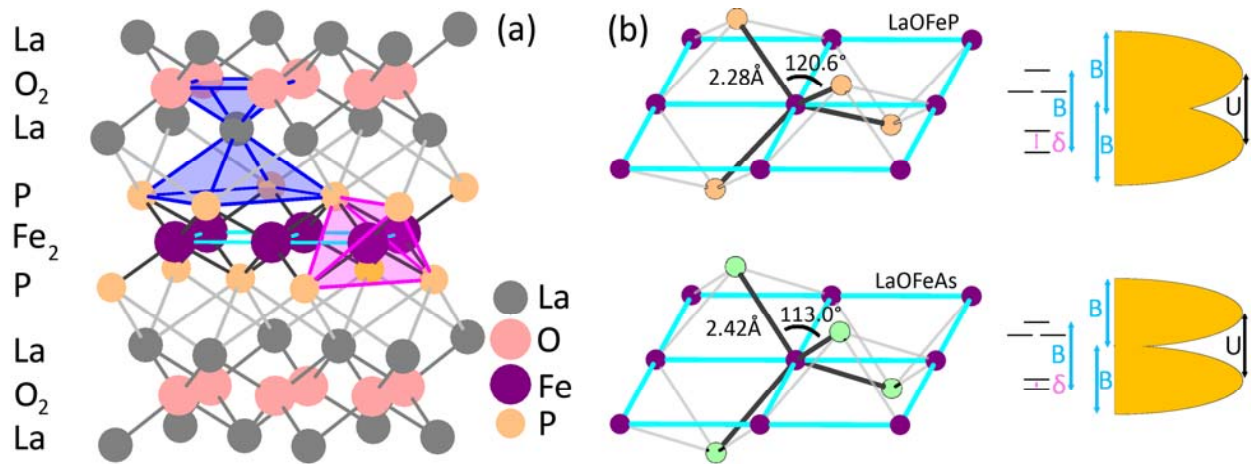

Figure 2

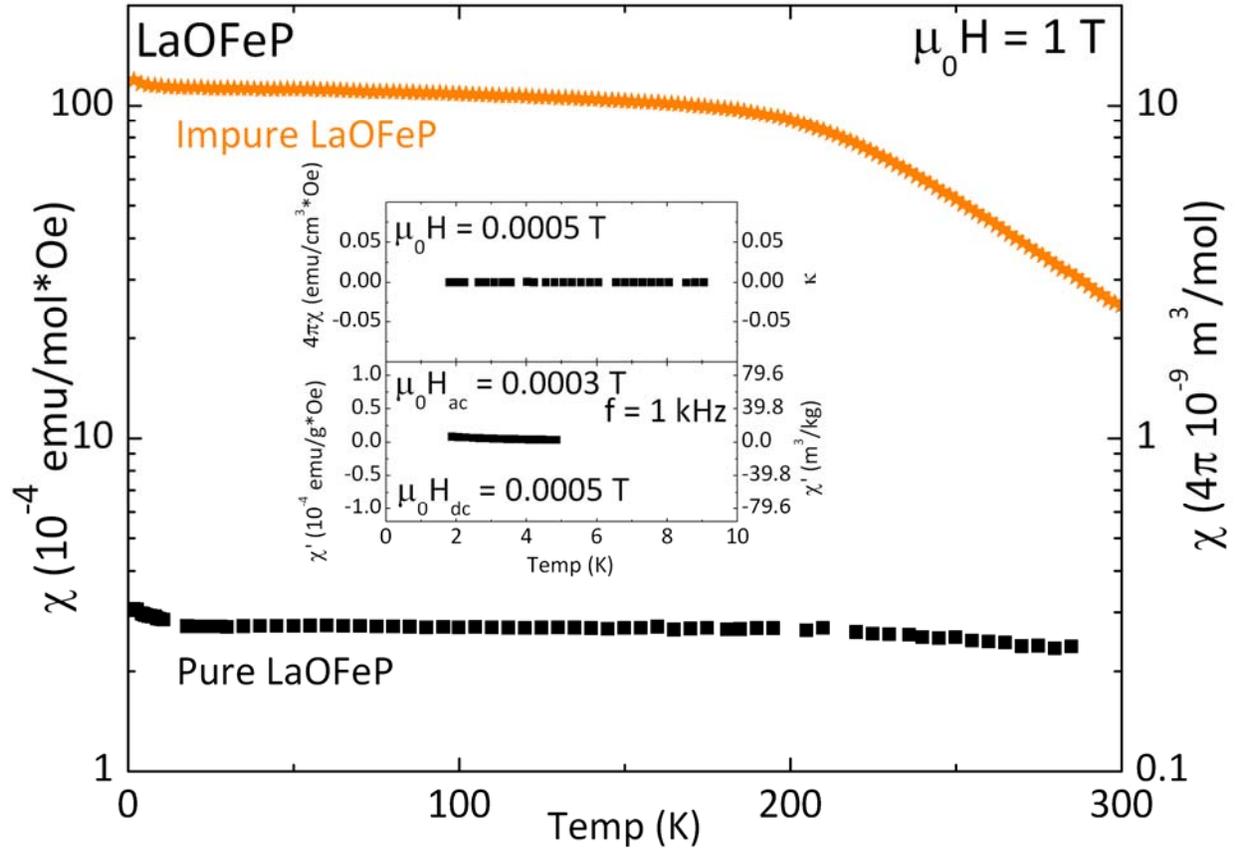

Figure 3

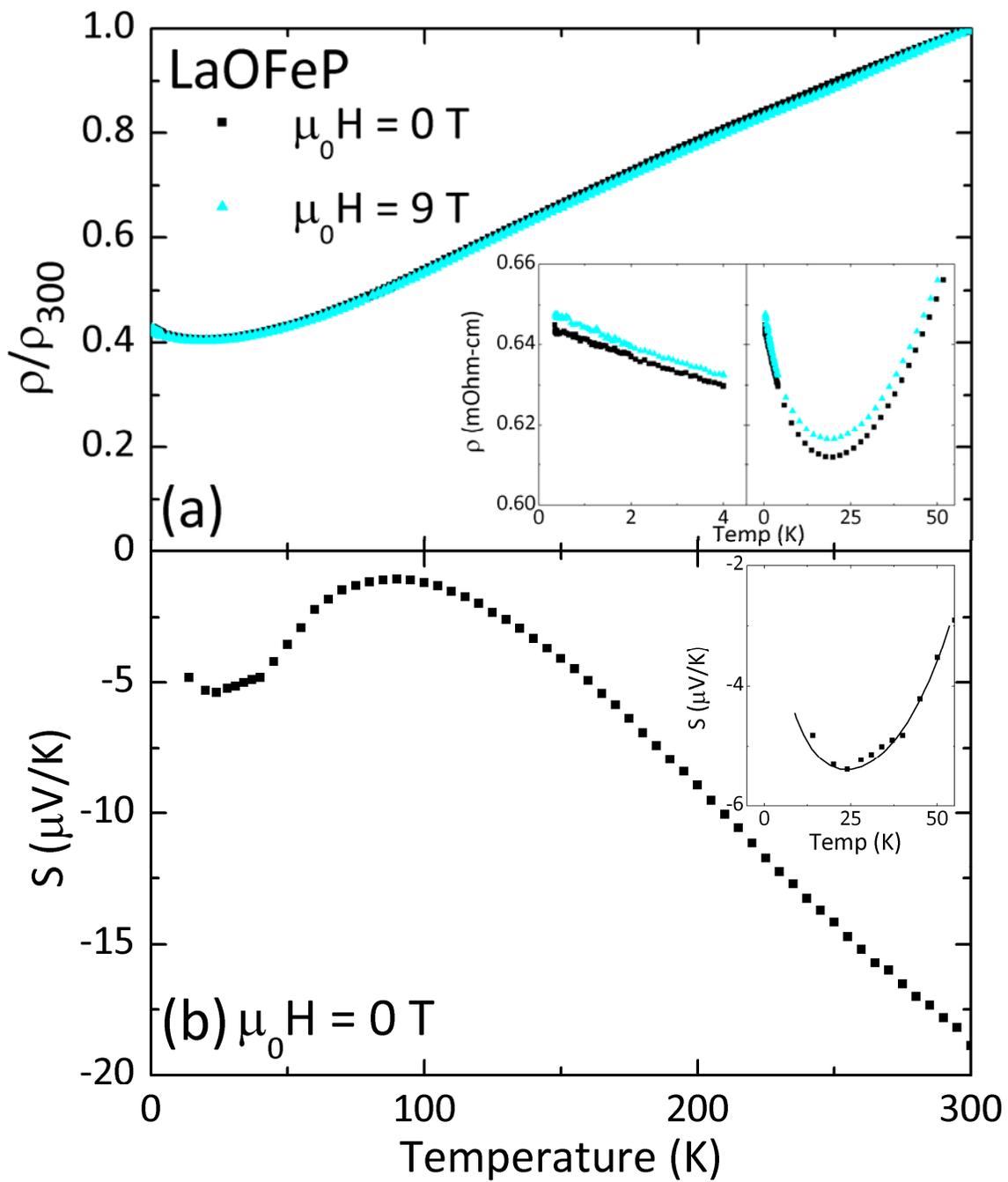

Figure 4

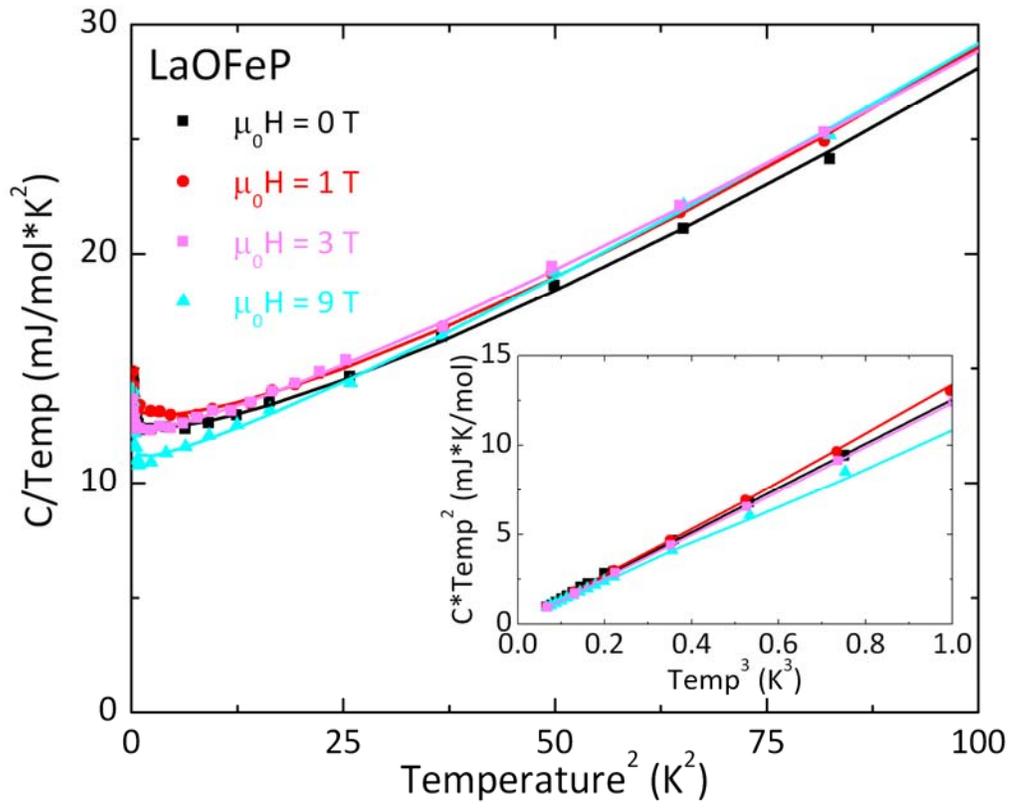

Figure 5

[37] $g(\varepsilon_F)_{calc} = 6 \frac{states}{eV \cdot cell}$, so $\gamma_{calc} = g(\varepsilon_F)_{calc} / 424.25 = 14.1 \frac{mJ}{mol \cdot K^2}$

[38] $\chi_{calc} = \mu_B^2 g(\varepsilon_F)_{calc} = 2.4 \cdot 10^{-9} \frac{m^3}{mol}$

[39] $\chi_{meas} = \chi_{pauli} + \chi_{landau} \approx \chi_{pauli} - \frac{1}{3}\chi_{pauli} = \frac{2}{3}\chi_{pauli}$, where the relation $\chi_{landau} = -\frac{1}{3}\chi_{pauli}$, assuming a free electron gas, has been used.

[40] For spin fluctuations, $\beta = \beta_3 - A \ln T_{sf}$ and thus a plot of $\beta$ versus $A$ will be linear with intercept $\beta_3$ and slope $-\ln T_{sf}$.

[41] For spin fluctuations, $A = \frac{\alpha \gamma_0}{T_{sf}^2}$, where $\gamma_0$ does not include any mass enhancement from the spin fluctuations and $\alpha = \frac{6\pi^2}{5}\frac{(S-1)^2}{S}$. For a system with negligible mass enhancement, $\gamma_0 = \gamma_{meas}$.